\documentclass[twocolumn,showpacs,preprintnumbers,amsmath,amssymb]{revtex4}
\usepackage{tabularx,graphicx}

\usepackage{color}
\usepackage{hyperref}
\hypersetup{
    colorlinks=true,
    linkcolor=blue,
    filecolor=blue,      
    urlcolor=blue,
}

\usepackage{color}

\usepackage{ulem}   

\begin{document}

\newcommand{\beq}{\begin{equation}}
\newcommand{\eeq}{\end{equation}}
\newcommand{\beqn}{\begin{eqnarray}}
\newcommand{\eeqn}{\end{eqnarray}}
\newcommand{\bmath}{\begin{subequations}}
\newcommand{\emath}{\end{subequations}}
\newcommand{\bra}[1]{\langle #1|}
\newcommand{\ket}[1]{|#1\rangle}

\title{Absence of high temperature superconductivity in hydrides under pressure}

\author{J. E. Hirsch$^{a}$  and F. Marsiglio$^{b}$ }
\address{$^{a}$Department of Physics, University of California, San Diego,
La Jolla, CA 92093-0319\\
$^{b}$Department of Physics, University of Alberta, Edmonton,
Alberta, Canada T6G 2E1}

\begin{abstract} 
The long-sought goal of room temperature superconductivity has reportedly recently been realized in a carbonaceous sulfur
hydride compound under high pressure, as reported by Snider et al \cite{roomt}. The evidence presented in that paper
is stronger than in other similar recent reports of high temperature superconductivity in hydrides under high
pressure \cite{eremetslah,hemleylah,eremetssh,yttrium,thorium,550}, and has been received with universal acclaim \cite{press1,press2,press3}.
Here we point out  that features of the experimental data shown in Ref.~\cite{roomt} indicate that the phenomenon observed in that
material is not
superconductivity. This observation calls into question  earlier similar claims of high temperature conventional 
superconductivity in hydrides under high pressure based on similar or weaker evidence
 \cite{eremetslah,hemleylah,eremetssh,yttrium,thorium,550}.

\end{abstract}
\maketitle 

In the Snider et al. paper  \cite{roomt} it is claimed  that the material is a weakly type II superconductor, 
with Ginzburg-Landau (GL) parameter $\kappa=\lambda(T)/\xi(T)\sim 1.5$, with $\lambda$ and $\xi$ the 
London penetration depth and coherence length respectively. This claim is based on an incorrect analysis.
From the experimental data close to $T_c$ an upper critical field $H_{c2}(T=0)=61.8T$ is inferred. This determines the
coherence length $\xi$  from the GL formula
\beq
H_{c2}=\frac{\phi_0}{2\pi \xi^2} .
\eeq
as $\xi=2.3nm$  \cite{roomt}, with $\phi_0=2.07\times 10^{-7}G- cm^2$ the flux quantum. 
In the figure caption of Extended Data Fig. 3, the authors state that they extract the penetration depth 
$\lambda(0)$ from
the formula
\beq
\lambda(0)=\frac{\phi_0}{2\sqrt{2}\pi H_c(0)\xi(0)}
\eeq
and state that $H_c(0)=61.8T$. However, as stated above, $61.8T$ is the upper critical field $H_{c2}$ inferred 
from experiment and not the thermodynamic critical field $H_c$ that enters in Eq. (2). Eq. (2) yields
$\lambda(0)=1.6nm$ (the quoted value of $3.8nm$ is a typo according to the authors \cite{typo}). 
Both values are incorrect.

The reality is that from the experimental results presented in the paper there is no way to extract
information on the value of the  penetration depth, so that both the numerical value of $\lambda(0)$ and its temperature dependence
presented in Extended Data Fig. 3 (b) and (c) are fiction, not reality. And the material is $not$ a weakly type II
superconductor, with GL parameter $\kappa \sim 1.5$ \cite{roomt}, as inferred by the authors \cite{remark1}.

Quite the contrary: if this material is a superconductor, it  should  be strongly type II, like
other superconductors with short coherence length like the 
high temperature cuprate superconductors or magnesium diboride ($MgB_2$). The coherence length inferred from the critical field data is
very short, $2.3nm$. Within the conventional theory of superconductivity \cite{tinkham},
which high temperature superconducting hydrides purportedly obey \cite{convth}, one can estimate the Fermi velocity $v_F$ from the
relation $\xi_0=\hbar v_F/(\pi \Delta(0))$, and the London penetration depth from
$\lambda(0)=\sqrt{m_ec^2/(4\pi n_s e^2})$ with $n_s$ the superfluid density \cite{tinkham}. Assuming for simplicity a spherical Fermi surface 
and the value of 
$\Delta(0)\sim 42 meV$ inferred from the measured $T_c$ \cite{roomt} yields  an estimated $\lambda(0)=113nm$. In reality,
the disorder present in these samples is likely to strongly reduce the value of the superfluid density 
$n_s$ increasing the magnitude of the penetration depth \cite{tinkham}. So we argue that
a GL ratio of $\kappa=\lambda(0)/\xi(0)\sim 113nm/2.3nm\sim 50$ is likely to be a lower bound for these materials.
For comparison, in the cuprate superconductors one has typically $\xi \sim 1.8nm$ and $\lambda\sim 180nm$, hence
$\kappa \sim 100$, and in $MgB_2$ $\xi \sim 5nm$ and $\lambda \sim 140nm$, hence $\kappa \sim 28$. 

The curves of resistance versus temperature shown in Fig. 1 show a remarkably sharp drop for all pressures shown. 
It is difficult to discern any transition width from the graph; we estimate it to be certainly less than 1K, as the paper also
states,
which corresponds to a fractional width of less than $0.5\%$. This is a remarkably sharp transition which is
rarely seen in any superconductor except for exceptionally pure single crystal samples of type I superconductors. The system under consideration here
is certainly not a single crystal and there are probably a range of compositions within the pressure cell.  The authors indicate that there are pressure gradients in the system to account for
observed differences in transition temperatures measured by resistivity and susceptibility. We argue that this is
inconsistent with the exceptionally sharp transitions displayed in Fig. 1.

This anomalous behavior becomes even clearer when considering Fig. 2. 
In type II superconductors  the resistive superconducting transition will necessarily be broadened in an applied
magnetic field. This is because the material enters into the mixed phase when the 
temperature is lowered or the magnetic field is decreased so that the applied magnetic field $H$ becomes smaller than 
$H_{c2}(T)$.
The magnetic field penetrates the material in the form of vortices carrying one flux quantum each,
with the number of vortices an increasing function of the applied field.
In the mixed phase a circulating current causes motion of vortices that dissipates energy so the resistivity is non-zero.
In the cuprates the broadening of the resistive transition is very large \cite{cuprate,cuprate2}, but it is a universal phenomenon
for all type II superconductors, becoming more pronounced the more strongly type II the material is and the higher
the  temperature is. Yet there is no indication in
Fig. 2b of $any$ broadening of the resistive transition under application of a magnetic field.

For quantitative comparison let us consider $MgB_2$, which, just like the hydrides under pressure, is universally believed to be a conventional superconductor. We also expect $MgB_2$ to be less strongly type II (smaller $\kappa$) than the material in the paper under 
consideration, as discussed above. Data for resistive transition in a field are given in ref. \cite{canfield},
Fig. 2. The resistive transition becomes   increasingly broader as the magnetic field increases, as expected,
contrary to what is seen in Fig. 2b of the paper of Snider et al \cite{roomt}. 
The upper critical field for $MgB_2$ is approximately $H_{c2}(0)=16T$. For an applied field 
$H=2.5T$, hence $H/H_{c2}(0)=0.15$, the transition is broadened over a range $\Delta T_c\sim 2.5 K$,
for critical temperature $T_c\sim 30K$, as seen in Fig. 2 of Ref.~\cite{canfield}. Therefore, for $MgB_2$
\beq
\frac{\Delta T_c}{T_c}(H/H_{c2}=0.15)\sim8.3\%.
\eeq

In contrast, consider the resistive transition shown in Fig. 2b of the paper under consideration  \cite{roomt} for applied magnetic
field $9T$. With estimated upper critical field $H_{c2}(0)=61.8 T$ this corresponds to also a ratio
$H/H_{c2}(0)\sim 0.15$.  From the data 
shown in Fig. 2b  we infer that the broadening is $\Delta T_c\sim 0.4K$, for $T_c\sim 265K$, hence 
\beq
\frac{\Delta T_c}{T_c}(H/H_{c2}=0.15)=0.15\%
\eeq
for carbonaceous sulfur
hydride under pressure.

Therefore, the broadening of the resistive transition in the case under consideration here  is
{\it at least a factor of 50 smaller than expected}. We say at least because we argued that this material
is likely to be more strongly type II than $MgB_2$, and the transition is at a much higher temperature.

Consideration of other type II superconductors, both materials considered to be conventional \cite{chevrel,cava,borocarbide,diamond}
and unconventional \cite{cuprate,cuprate2,organic,pnictide},
confirm our argument: broadening of the resistive transition in a field is a universal property of type II superconductors.
We are not aware of a single example in the scientific literature showing a resistive transition in magnetic field
of any type II superconductor that would be nearly as sharp as shown in Ref.~\cite{roomt}, or that would show 
no broadening in a magnetic field as this material shows.
The transition between normal and superconducting states in type II conventional superconductors
and resulting transport properties  are well understood   \cite{fluxflow,fluxflow2}, flux-flow resistivity arises from 
motion of vortices in the mixed phase broadening the transition. At the high temperatures where the transition apparently occurs
in this material thermal activation of vortices should be pronounced and the role of pinning centers reduced, increasing
the broadening.
Broadening of the resistive transition is also quite well understood for some unconventional superconductors like the cuprates,
where some novel physics and additional dissipation mechanisms may exist  \cite{fisher}. In all cases, both conventional and unconventional superconductors,
a broadening of the resistive transition in a magnetic field universally occurs. It is an inescapable consequence of the physics
of superconductors where the coherence length is substantially shorter than the London penetration depth
so that the system first enters a mixed state,  
independently of the particular mechanism giving rise to superconductivity.

Because the data for the resistive transition in a field of this material diverge from this established knowledge by
a factor of at least 50 as shown above, we conclude that it is impossible that the data  in Fig. 2b of this paper 
 reflect a transition to a superconducting state.

What could be the origin of the extremely sharp transitions seen in Fig. 2b?  Most likely, a metallic conduction path was
suddenly established where previously there was none. Note also that all the resistance curves shown in Ref. \cite{roomt} are obtained during  {\it warming} cycles.  So what the curves are showing is not a ``sharp drop in resistance'', as the paper   claims,
but instead a sharp resistance {\it increase}. 
 The material is inhomogeneous and likely to be 
composed of metallic and non-metallic regions, and at some given pressure and temperature a metallic path that existed
between electrodes at a lower temperature or higher pressure may be cut off, suddenly increasing the resistance. 
The transition can be extremely sharp in that case, with the higher pressure or lower temperature state having much lower
electrical resistance than the lower pressure/higher temperature state. But it  would not be a superconducting state.
Note that  in the example shown in Extended Data Fig. 4 the low temperature
resistance is clearly not zero.

If indeed the curve shown in Fig. 2 for magnetic field $9T$ does not reflect  a transition between a 
normal and a  superconducting state, 
and we can't conceive of any way that this could not be so, it is
necessarily due to different physics. This same different physics is likely to be responsible for the similar behavior
shown in all the other resistivity curves in this paper, hence none of them would provide evidence for superconductivity either.
Furthermore  this calls into question that  the weaker similar evidence seen in other papers on hydrides in recent years
  \cite{eremetslah,hemleylah,eremetssh,yttrium,thorium,550}
is evidence of superconductivity. Rather, all these experimental results that are so difficult to reproduce 
have likely a common origin that still needs to be elucidated, but it is not superconductivity.

In conclusion, unless or until the exceptional sharpness of the resistive transitions shown in this paper can be plausibly explained within the
generally accepted conventional theory of superconductivity, the experimental evidence on high temperature 
hydride superconductivity that exists so far \cite{review2,roomt} remains questionable.

\begin{acknowledgments}
We acknowledge clarifying correspondence with the authors of Ref.  \cite{roomt}. 
FM 
was supported in part by the Natural Sciences and Engineering
Research Council of Canada (NSERC).  

\end{acknowledgments}

\noindent {\bf Author contributions:} JEH and FM contributed equally to the work.

\noindent {\bf Competing interests:} the authors declare no competing interests.


\begin{references}
\bibitem{remark1} The rest of Extended Data Fig. 3 is entirely theoretical, as far as we can tell, and parts of it are misapplied. 
For example, part (d)
     must have a typo in the caption, and the curves shown represent some sort of expansion in temperature near $T_c$ (though not the one from BCS), and yet the (theoretical) plot extends over all temperatures down to zero temperature.
\end{references}
 \end{document}